\PassOptionsToPackage{unicode}{hyperref}
\PassOptionsToPackage{hyphens}{url}
\documentclass[
]{article}
\usepackage{amsmath,amssymb}
\usepackage{lmodern}
\usepackage{iftex}
\ifPDFTeX
  \usepackage[T1]{fontenc}
  \usepackage[utf8]{inputenc}
  \usepackage{textcomp} 
\else 
  \usepackage{unicode-math}
  \defaultfontfeatures{Scale=MatchLowercase}
  \defaultfontfeatures[\rmfamily]{Ligatures=TeX,Scale=1}
\fi
\IfFileExists{upquote.sty}{\usepackage{upquote}}{}
\IfFileExists{microtype.sty}{
  \usepackage[]{microtype}
  \UseMicrotypeSet[protrusion]{basicmath} 
}{}
\makeatletter
\@ifundefined{KOMAClassName}{
  \IfFileExists{parskip.sty}{%
    \usepackage{parskip}
  }{
    \setlength{\parindent}{0pt}
    \setlength{\parskip}{6pt plus 2pt minus 1pt}}
}{
  \KOMAoptions{parskip=half}}
\makeatother
\usepackage{xcolor}
\IfFileExists{xurl.sty}{\usepackage{xurl}}{} 
\IfFileExists{bookmark.sty}{\usepackage{bookmark}}{\usepackage{hyperref}}
\hypersetup{
  pdftitle={Computational philosophy of science},
  pdfauthor={Michał J. Gajda},
  hidelinks,
  pdfcreator={LaTeX via pandoc}}
\providecommand{\tightlist}{%
  \setlength{\itemsep}{0pt}\setlength{\parskip}{0pt}}
\newlength{\cslhangindent}
\newlength{\csllabelwidth}
\newlength{\cslentryspacingunit} 
\newenvironment{CSLReferences}[2] 
 {
  \setlength{\parindent}{0pt}
  \ifodd #1
  \let\oldpar\par
  \def\par{\hangindent=\cslhangindent\oldpar}
  \fi
  \setlength{\parskip}{#2\cslentryspacingunit}
 }%
 {}
\usepackage{calc}

\ifLuaTeX
  \usepackage{selnolig}  
\fi

\title{Computational philosophy of science}
\usepackage{etoolbox}
\makeatletter
\providecommand{\subtitle}[1]{
  \apptocmd{\@title}{\par {\large #1 \par}}{}{}
}
\makeatother
\subtitle{Short paper}
\author{Michał J. Gajda}
\date{}

\begin{document}
\maketitle
\begin{abstract}
Philosophy of science attempts to describe all parts of the scientific
process in a general way in order to facilitate the description,
execution and improvements of this process.

So far, all proposed philosophies have only covered existing processes
and disciplines partially and imperfectly. In particular logical
approaches have always received a lot of attention due to attempts to
fundamentally address issues with the definition of science as a
discipline with reductionist theories.

We propose a new way to approach the problem from the perspective of
computational complexity and argue why this approach may be better than
previous propositions based on pure logic and mathematics.
\end{abstract}

Many scientists call for distinguishing computational science as a new
kind of science (Wolfram 2002; Angius, Primiero, and Turner 2021). It is
extremely important to provide a sensible philosophical definition of
computational science since the discipline is rooted in disciplines of
data science and empirical modelling of nature. The search for a
satisfactory definition leads us to a new approach to the philosophy of
science. It suggests that computational science may not be a new formal
method of program construction but also a new approach to the philosophy
of science itself. And indeed allows us to generalization of philosophy
of science in a much more satisfactory way than many efforts to
date(Kuhn 1970; Rodych 2018; Laplace 1902; Hoang 2020).

\hypertarget{introduction}{%
\section{Introduction}\label{introduction}}

It has been argued that the development of computational capability was
a significant force in human history. One can start with ancient Egypt,
where mathematics and engineering were reserved to priests, through
geographic discoveries, when accessible astronomy and precise
computation allowed for daily course corrections, til now, when the
availability of personal computing allows nearly all scientists tackle
much greater challenges through bioinformatics, galaxy simulations, or
mass analysis of social networks (Himelboim 2017).

In the current philosophy of science, it is argued that the mass of
empirical data allow us quicker to falsify less useful scientific
theories, and thus that overall quality of our base theories has risen
exponentially. However, at the same time, it is argued that despite more
advanced methodologies, basic facts published by scientific journals
have a higher and higher probability of being false (Ioannidis 2005;
Economist 2013; Colquhoun 2017).

Indeed it seems that philosophers of science have up to now failed to
reconcile our feeling of growth of scientific knowledge, with the
statistical measures proposed to quantify it.

\hypertarget{definitions}{%
\section{Definitions}\label{definitions}}

\emph{Empiricism} is a philosophical view that all concepts originate
from experience, and \emph{empirical science} is thus science is thus
knowledge based on experimental confirmation and falsification. Thus
scientific theories are those that may be subject to continued revision
based on experiments.

\emph{Falsificationism} claims that most statements cannot be logically
deduced but only disproved by contrary evidence. Thus scientific
knowledge develops by \emph{falsification} of theories by experience.
Knowledge may be scientific only if it is falsifiable. Failed attempts
at falsification increase our confidence in theories, whereas successful
falsifications drive us to develop new theories(Popper 1934).

\emph{Bayesianism} claims that all knowledge is subject to varying
degrees of probability, and thus updating our beliefs should be subject
to Bayes' rule(Laplace 1902; Hoang 2020).

\emph{Predictivism} introduced in this work claims that we need to go
further than just the requirement of empirical validation and a
probabilistic view of knowledge. All scientific knowledge is just a way
to provide probabilistic predictions and is subject to constraints of
our computational abilities. Thus theories that have been falsified, but
are computationally inexpensive, should be still considered valid
scientific knowledge since they still provide predictions at lower costs
than more complex theories.

\hypertarget{research-questions}{%
\section{Research questions}\label{research-questions}}

We would like to discuss the following research questions:

\begin{description}
\tightlist
\item[RQ1]
What are traits of science as opposed to other human endeavours?
\item[RQ2]
Can we define science in such a way that it encompasses all current
activities?
\item[RQ3]
How do the philosophies of science differ in their applicability to
different domains and scientific activities?
\end{description}

\hypertarget{proposal}{%
\section{Proposal}\label{proposal}}

We propose to generalize the naive principle of simplification and
computation, that says that theories allowing us for simpler
explanations are indeed better scientific theories.

And given the logical development of the theory of falsification, we
want to eliminate absurdity of negative criterion for theoretical
validity.

After all, scientists willingly use falsified theories as tools when no
better theories have been created. It is rare to argue that these
falsified theories are a step back when they have some prediction.

Indeed we claim that a series of empirical theories may be all somewhat
false due to overgeneralization, and they still could represent the
growth of scientific knowledge, if each consecutive theory provides more
accurate predictions.

Thus we claim that the \emph{predictive power} should be the most
important quality of science. That makes science a practical tool that
allows us to simplify and predict a universe of facts and explains the
popularity of already falsified theories. Simpler theory may have
simpler computational requirements and thus make predictions \emph{more
accessible} for the scientists. Thus falsified theories become usual
stepping stones in the hierarchy of theories, and an essential part of
the holistic view of science as a predictive tool.

One could be even tempted to draw theories in such a hierarchy and
conclude that for a plot where the X-axis would be computational
complexity of prediction, and Y-axis would be the accuracy of the
prediction, useful scientific theories would go all the way along the
waterfront of preference function.

\hypertarget{scientific-theories-and-methodologies-considered}{%
\section{Scientific theories and methodologies
considered}\label{scientific-theories-and-methodologies-considered}}

Here we give a few examples of how the framework of predictivism may
explain the evolution of science.

\hypertarget{newtonian-dynamics-vs-relativity-theory}{%
\subsection{Newtonian dynamics vs relativity
theory}\label{newtonian-dynamics-vs-relativity-theory}}

Newtonian dynamics(Newton 1687) is a body of theories derived from
Newton's three laws of motion. Its mathematical formulations are the
basis for a significant body of state-of-art physical descriptions in
physics, engineering, and architecture. However, it has since been
falsified by experiments confirming the speed light limit and replaced
by general relativity theory(Einstein 1905) and postnewtonian
approximations(Chandrasekhar 1969).

The discovery of quantum mechanics also forced a revision of the Newton
equation(Meter 2011).

Why is the theory still used long after its falsification? Would it not
be prudent to stop teaching an obviously false theory and convert all
mathematical descriptions to general relativity?

While it seems possible for macroscopic interactions, it would also
significantly increase the complexity of these descriptions.

Indeed, practitioners in this field have long argued that sound research
is characterized by the interaction of falsifiable theories,
experimental measurements, and effective approximations(Hartmann 2001).

Also, such replacement is not known in generality for quantum mechanics.

\hypertarget{quantum-mechanics-vs-general-relativity}{%
\subsection{Quantum mechanics vs general
relativity}\label{quantum-mechanics-vs-general-relativity}}

World description by quantum mechanics(Feynman 1950; Schweber 1986) is
considered inconsistent with that of general relativity. Whereas general
relativity correctly describes phenomena at the macroscale, quantum
mechanics correctly describes sub-nanometer scale processes.

Thus we see that both theories were falsified when tried in full
generality!

\hypertarget{biomolecular-structure-prediction}{%
\subsection{Biomolecular structure
prediction}\label{biomolecular-structure-prediction}}

Having developed theories of protein structure(Pauling, Corey, and
Branson 1951), scientists attempted to describe the mechanics of small
and large molecules by means of molecular mechanics(Allinger 1959). This
theory uses Newtonian laws of motion, together with a simplified view of
chemical potentials, to describe forces acting on different atoms and
thus model their interaction.

This approach was so successful that nowadays, experimentalists turn to
molecular mechanics to solve complex experimental data(Joosten et al.
2011). This is an example of a theory that is used despite its
inaccuracy as compared with quantum chemistry (Warshel and Levitt 1976).
Indeed its success has led to further simplifications: approximating
covalent bonds by the stiff connections of fixed length, and recomputing
all forces in angle space has given birth to torsion angle mechanics(Bae
and Haug 1987; Kumar, Hosur, and Govil 1991). This method has sped up
protein structure determination from days and hours to minutes for
computing the simple model(Winn et al. 2002). Even further is the
Rosetta method(Rohl et al. 2004), which uses molecular mechanics for the
last stage of structure polishing but starts by simply trying to copy
and paste fragments of existing protein structures from Protein
Databank(H. M. Berman 2003, 2007; Consortium 2019) database and checking
for geometrical compatibility\footnote{Called ``steric clashes''.}.

This is an excellent example of a modern scientific domain that has
developed by going towards computationally more straightforward and
faster methods, in order to achieve overall improvements in the number
and accuracy of predictions.

This also shows how great progress in science is achieved by going from
complex to simpler explanations, even though the simpler explanations
are less accurate and indeed more easily falsified(Rauscher et al. 2015)

\hypertarget{dna-barcoding-vs-computational-phylogeny}{%
\subsection{DNA barcoding vs computational
phylogeny}\label{dna-barcoding-vs-computational-phylogeny}}

Add DNA barcoding as a recently proposed approximation method that tries
to reduce the computational load of getting the right answer.

DNA barcoding(Woese, Kandler, and Wheelis 1990; Hebert et al. 2003) is a
recent method of species identification that attempts to decrease the
computational complexity of the process. It is based on a comparison of
gene regions with a less interspecific variation. Alternative ``golden
standard of truth'' for species differentiation nowadays is
computational phylogeny analysis, but his method requires extensive
sequencing of the genome and computational analysis that requires
computational resources as well as expertise. That is why DNA barcoding
was proposed as a method that can be used by the wider society of
scientists. DNA barcoding is also faster than traditional morphological
comparison methods and can be used when fruits and pollens are
unavailable.

\hypertarget{history}{%
\subsection{History}\label{history}}

Science of history can also be described as following from the known
\emph{historical records} to infer likely \emph{historical
facts}\footnote{The historical events may not have been exactly
  correspond to the events in records, and this is an important part of
  the process} in order to describe processes that shaped human history.

This seems hard to describe in either a classical framework of seeking
the truth (since we know that a significant percentage of the facts will
be untrue), within the framework of empiricism (since we can hardly
\emph{repeat the experiment}) and within the framework of
falsificationism (since strict falsification of most historical facts
would require time travel).

However, it is easy to describe within a framework of predictivism: we
say that inferred historical events are most likely, and the simplest
explanation of the existing historical records. In the same manner, we
discuss historical forces and processes: they are more general, shorter,
and thus more predictive explanations of the historical facts than
enumerating the facts themselves.

This shows how the framework of predictivism can describe science
outside of the empirical domain.

\hypertarget{philosophy}{%
\subsection{Philosophy}\label{philosophy}}

In our proposal, some fields of philosophical endeavour can be rephrased
into fields of predictive science:

\begin{itemize}
\tightlist
\item
  ontology as a way of predicting mutual relations of abstract terms and
  classifications
\item
  axiology as a way of predicting consistency of value judgements
\end{itemize}

In this sense, we can straightforwardly state that philosophy is indeed
\emph{empirical} science since we have clear metrics of success for each
theory.

\hypertarget{engineering-as-the-science-of-achieving-goals}{%
\subsection{Engineering as the science of achieving
goals}\label{engineering-as-the-science-of-achieving-goals}}

Similarly, many engineering and applied mechanics domains are obviously
scientific: for example, software engineering is the art of predicting
what software systems will satisfy design (or business) goals while
optimizing the work effort.

We can also argue that such a definition of science makes it easy to
explain improved fitness achieved by individuals and organizations using
scientific methods to compete. Even if these methods are based on
falsifiable theories, predictive power allows these entities to improve
their fitness by either reducing computational effort needed for
situational awareness or providing increased situational awareness.

\hypertarget{what-is-the-value-of-the-theory}{%
\section{What is the value of the
theory?}\label{what-is-the-value-of-the-theory}}

What differentiates science from epistemic decisions of the individual
are that scientific theories are transmitted within the community, and
verifiable by community members.

The first factor allows us to build scientific theories across the chasm
of time and to make steady progress in our understanding of the world.
But the second factor is also important: it allows us to ``discard
spam'': theories contributing less than the others, by the merit of
providing less truthy or more computationally expensive predictions.
This allows us to avoid an explosion of untruths that would otherwise
accumulate in unfiltered culture. We still have a ``valuation'' of
scientific theories. One such valuation is truth-untruth; however,
Laplace posited that we could only assess the probability of the truth,
so bayesian interpretation by the probability of truth is in order.
Additional gradation may be that of Kolmogorov complexity of the
explanation since simpler explanations are preferred.

The philosophy described here points out that besides the probability of
truthiness, we also have a computational complexity of using the theory
in practice. Epistemological preference would thus only choose theories
that either enjoy a higher probability of truthiness OR lower
computational complexity. This is important both for humans (who can use
multiple methods but see only a few moves ahead in games like chess) and
for computers. One can, of course, argue that epistemological
computational cost is different for humans and for computers. They have
different computational models: humans have massively parallel neural
networks of only limited depth and a limited number of layers that can
be used during the time for computation; computers have highly
sequential CPUs that allow for billions of operations per second but
only limited parallelism. One can also consider the cost of human
training or implementation of prediction software on the computing
platform, but the resolution will be similar: the overall computational
complexity of both implementing and running the predictive theory will
dictate its value much more than its truthiness. Because most of the
time, we only predict as much as our computational power predicts, not
only saving energy this way but providing a timely world-view to our
decision making.

\hypertarget{family-of-optimal-theories}{%
\section{Family of optimal theories}\label{family-of-optimal-theories}}

It also allows us to talk about ultrafinitist logic, calculus of
constructions, intuitionist logic, and classical logic on equal footing
by considering them some possible \emph{optimal} points on the scale
between the accessibility of logic and the desire for the elimination of
all paradoxes. Thus one could argue that the recent development of
multimodal and other variant logics is not experimentation, but a
natural exploration of \emph{all possible optimal theories} within this
space.

\hypertarget{discussion}{%
\section{Discussion}\label{discussion}}

We see how predictivism provides an accessible basis for defining
various domains of science, and explains the continued development of
readily falsified theories. We also propose a metric that allows us to
compare theories on a non-bibliometric basis, and thus judge whether
they are optimal in the sense of providing the best prediction for a
given computational complexity or not.

We thus proposed a re-evaluation of the philosophy of science through
the lens of computational complexity. This suggests that computational
complexity may be indeed a fundamental concept for the understanding of
science in general, just like the theory of probability is.

\hypertarget{bibliography}{%
\section{Bibliography}\label{bibliography}}

\hypertarget{refs}{}
\begin{CSLReferences}{1}{0}
\leavevmode\vadjust pre{\hypertarget{ref-molecular-mechanics}{}}%
Allinger, Norman. 1959. {``Conformational Analysis. III. Applications to
Some Medium Ring Compounds1,2.''} \emph{Journal of the American Chemical
Society} 81 (November). \url{https://doi.org/10.1021/ja01530a049}.

\leavevmode\vadjust pre{\hypertarget{ref-sep-phil-cs}{}}%
Angius, Nicola, Giuseppe Primiero, and Raymond Turner. 2021. {``{The
Philosophy of Computer Science}.''} In \emph{The {Stanford} Encyclopedia
of Philosophy}, edited by Edward N. Zalta, Spring 2021.
\url{https://plato.stanford.edu/archives/spr2021/entries/computer-science/};
Metaphysics Research Lab, Stanford University.

\leavevmode\vadjust pre{\hypertarget{ref-torsion-angle-dynamics-mechanical}{}}%
Bae, Dae-Sung, and Edward J. Haug. 1987. {``A Recursive Formulation for
Constrained Mechanical System Dynamics: Part i. Open Loop Systems.''}
\emph{Mechanics of Structures and Machines} 15 (3): 359--82.
\url{https://doi.org/10.1080/08905458708905124}.

\leavevmode\vadjust pre{\hypertarget{ref-postnewtonianApproximations}{}}%
Chandrasekhar, S. 1969. {``{Conservation Laws in General Relativity and
in the Post-Newtonian Approximations}''} 158 (October): 45.
\url{https://doi.org/10.1086/150170}.

\leavevmode\vadjust pre{\hypertarget{ref-p-values}{}}%
Colquhoun, David. 2017. {``The Reproducibility of Research and the
Misinterpretation of p-Values.''} \emph{Royal Society Open Science} 4
(12): 171085--85. \url{https://doi.org/10.1098/rsos.171085}.

\leavevmode\vadjust pre{\hypertarget{ref-PDB3}{}}%
Consortium, wwPDB. 2019. {``Protein Data Bank: The Single Global Archive
for 3d Macromolecular Structure Data.''} \emph{Nucleic Acids Research}
47 (12): 520--28. \url{https://doi.org/10.1093/nar/gky949}.

\leavevmode\vadjust pre{\hypertarget{ref-trouble-in-the-lab}{}}%
Economist. 2013. 2013.
\url{https://www.economist.com/briefing/2013/10/18/trouble-at-the-lab}.

\leavevmode\vadjust pre{\hypertarget{ref-relativity-theory}{}}%
Einstein, A. 1905. {``Zur Elektrodynamik Bewegter Körper.''}
\emph{Annalen Der Physik} 322 (10): 891--921.
https://doi.org/\url{https://doi.org/10.1002/andp.19053221004}.

\leavevmode\vadjust pre{\hypertarget{ref-quantumDynamics}{}}%
Feynman, R. P. 1950. {``Mathematical Formulation of the Quantum Theory
of Electromagnetic Interaction.''} \emph{Phys. Rev.} 80 (November):
440--57. \url{https://doi.org/10.1103/PhysRev.80.440}.

\leavevmode\vadjust pre{\hypertarget{ref-PDB1}{}}%
H. M. Berman, H. Nakamura, K. Henrick. 2003. {``Announcing the Worldwide
Protein Data Bank.''} \emph{Nature Structural Biology} 10 (12): 980.

\leavevmode\vadjust pre{\hypertarget{ref-PDB2}{}}%
---------. 2007. {``(2007) the Worldwide Protein Data Bank (wwPDB):
Ensuring a Single, Uniform Archive of PDB Data Nucleic Acids Res. 35
(Database Issue): D301-3.''} \emph{Nucleic Acids Research} 35 (Database
issue) (3): 301.

\leavevmode\vadjust pre{\hypertarget{ref-effectiveFieldTheories}{}}%
Hartmann, Stephan. 2001. {``Effective Field Theories, Reductionism and
Scientific Explanation.''} \emph{Studies in History and Philosophy of
Science Part B: Studies in History and Philosophy of Modern Physics} 32
(2): 267--304.
https://doi.org/\url{https://doi.org/10.1016/S1355-2198(01)00005-3}.

\leavevmode\vadjust pre{\hypertarget{ref-dna-barcoding-paper}{}}%
Hebert, Paul D. N., Alina Cywinska, Shelley L. Ball, and Jeremy R.
deWaard. 2003. {``Biological Identifications Through DNA Barcodes.''}
\emph{Proceedings of the Royal Society of London. Series B: Biological
Sciences} 270 (1512): 313--21.
\url{https://doi.org/10.1098/rspb.2002.2218}.

\leavevmode\vadjust pre{\hypertarget{ref-social-networks}{}}%
Himelboim, Itai. 2017. {``Social Network Analysis (Social Media).''} In.
\url{https://doi.org/10.1002/9781118901731.iecrm0236}.

\leavevmode\vadjust pre{\hypertarget{ref-hoang}{}}%
Hoang, Lê Nguyên. 2020. \emph{{The Equation of Knowledge: From Bayes'
Rule to a Unified Philosophy of Science}}. 1st ed. Chapman; Hall/CRC.
\url{https://doi.org/10.1201/9780367855307}.

\leavevmode\vadjust pre{\hypertarget{ref-ioannidis}{}}%
Ioannidis, John P. A. 2005. {``Why Most Published Research Findings Are
False.''} \emph{PLOS Medicine} 2 (8).
\url{https://doi.org/10.1371/journal.pmed.0020124}.

\leavevmode\vadjust pre{\hypertarget{ref-pdb-remodeling}{}}%
Joosten, Robbie P., Krista Joosten, Serge X. Cohen, Gert Vriend, and
Anastassis Perrakis. 2011. {``{Automatic rebuilding and optimization of
crystallographic structures in the Protein Data Bank}.''}
\emph{Bioinformatics} 27 (24): 3392--98.
\url{https://doi.org/10.1093/bioinformatics/btr590}.

\leavevmode\vadjust pre{\hypertarget{ref-Kuhn}{}}%
Kuhn, Thomas S. 1970. \emph{The Structure of Scientific Revolutions}.
Chicago: University of Chicago Press.

\leavevmode\vadjust pre{\hypertarget{ref-torsion-angle-dynamics}{}}%
Kumar, R. Ajay, R. V. Hosur, and G. Govil. 1991. {``Torsion Angle
Approach to Nucleic Acid Distance Geometry: TANDY.''} \emph{Journal of
Biomolecular NMR} 1 (4): 363--78.
\url{https://doi.org/10.1007/BF02192860}.

\leavevmode\vadjust pre{\hypertarget{ref-laplace}{}}%
Laplace, Pierre Simon. 1902. \emph{A Philosophical Essay on
Probabilities}. New York: J. Wiley \& Sons.

\leavevmode\vadjust pre{\hypertarget{ref-newton-schroedinger}{}}%
Meter, J R van. 2011. {``Schrödinger--Newton {`Collapse'} of the
Wavefunction.''} \emph{Classical and Quantum Gravity} 28 (21): 215013.
\url{https://doi.org/10.1088/0264-9381/28/21/215013}.

\leavevmode\vadjust pre{\hypertarget{ref-newton}{}}%
Newton, Isaac. 1687. \emph{Philosophiae Naturalis Principia
Mathematica}. William Dawson \& Sons Ltd., London.

\leavevmode\vadjust pre{\hypertarget{ref-protein-structure}{}}%
Pauling, Linus, Robert Corey, and H. Branson. 1951. {``The Structure of
Proteins: Two Hydrogen-Bonded Helical Configurations of the Polypeptide
Chain.''} \emph{Proceedings of The National Academy of Sciences - PNAS}
37 (April). \url{https://doi.org/10.1073/pnas.37.4.205}.

\leavevmode\vadjust pre{\hypertarget{ref-popper}{}}%
Popper, K. R. 1934. \emph{The Logic of Scientific Discovery}. London:
Hutchinson.

\leavevmode\vadjust pre{\hypertarget{ref-sarah-disordered-proteins}{}}%
Rauscher, Sarah, Vytautas Gapsys, Michal J. Gajda, Markus Zweckstetter,
Bert L. de Groot, and Helmut Grubmüller. 2015. {``Structural Ensembles
of Intrinsically Disordered Proteins Depend Strongly on Force Field: A
Comparison to Experiment.''} \emph{Journal of Chemical Theory and
Computation} 11 (11): 5513--24.
\url{https://doi.org/10.1021/acs.jctc.5b00736}.

\leavevmode\vadjust pre{\hypertarget{ref-Wittgenstein}{}}%
Rodych, Victor. 2018. {``{Wittgenstein's Philosophy of Mathematics}.''}
In \emph{The {Stanford} Encyclopedia of Philosophy}, edited by Edward N.
Zalta, Spring 2018.
\url{https://plato.stanford.edu/archives/spr2018/entries/wittgenstein-mathematics/};
Metaphysics Research Lab, Stanford University.
\url{https://plato.stanford.edu/archives/spr2018/entries/wittgenstein-mathematics/}.

\leavevmode\vadjust pre{\hypertarget{ref-rosetta}{}}%
Rohl, Carol A., Charlie E. M. Strauss, Kira M. S. Misura, and David
Baker. 2004. {``Protein Structure Prediction Using Rosetta.''} In
\emph{Numerical Computer Methods, Part d}, 383:66--93. Methods in
Enzymology. Academic Press.
https://doi.org/\url{https://doi.org/10.1016/S0076-6879(04)83004-0}.

\leavevmode\vadjust pre{\hypertarget{ref-feynmannHistory}{}}%
Schweber, Silvan S. 1986. {``Feynman and the Visualization of Space-Time
Processes.''} \emph{Rev. Mod. Phys.} 58 (April): 449--508.
\url{https://doi.org/10.1103/RevModPhys.58.449}.

\leavevmode\vadjust pre{\hypertarget{ref-quantum-molecular-mechanics}{}}%
Warshel, A., and M. Levitt. 1976. {``Theoretical Studies of Enzymic
Reactions: Dielectric, Electrostatic and Steric Stabilization of the
Carbonium Ion in the Reaction of Lysozyme.''} \emph{Journal of Molecular
Biology} 103 (2): 227--49.
https://doi.org/\url{https://doi.org/10.1016/0022-2836(76)90311-9}.

\leavevmode\vadjust pre{\hypertarget{ref-computational-protein-structure-determination}{}}%
Winn, M. D., A. W. Ashton, P. J. Briggs, C. C. Ballard, and P. Patel.
2002. {``{Ongoing developments in {\emph{CCP}}4 for high-throughput
structure determination}.''} \emph{Acta Crystallographica Section D} 58
(11): 1929--36. \url{https://doi.org/10.1107/S0907444902016116}.

\leavevmode\vadjust pre{\hypertarget{ref-dna-barcoding-proposal}{}}%
Woese, C R, O Kandler, and M L Wheelis. 1990. {``Towards a Natural
System of Organisms: Proposal for the Domains Archaea, Bacteria, and
Eucarya.''} \emph{Proceedings of the National Academy of Sciences} 87
(12): 4576--79. \url{https://doi.org/10.1073/pnas.87.12.4576}.

\leavevmode\vadjust pre{\hypertarget{ref-wolfram-new-kind-of-science}{}}%
Wolfram, Stephen. 2002. \emph{{A new kind of science}}. 1st ed. Wolfram
Media. \url{https://www.wolframscience.com/nks/}.

\end{CSLReferences}

\end{document}